\documentclass[12pt]{article}
\usepackage{amsfonts}
\usepackage{latexsym}
\usepackage{amsmath,amssymb}
\usepackage{verbatim}
\usepackage{graphicx}

\usepackage{setspace}

\usepackage[textheight=9in, textwidth=6.5in, letterpaper]{geometry}
\def\half{{1\over 2}}
\numberwithin{equation}{section}

\def\ip{${\mathcal I}^+$}

\def\e{{\epsilon}}

 \def\p{\partial}
 \def\bz{{\bar z}}
 
\def\0{{(0)}}
\def\1{{(1)}}
\def\2{{(2)}}

\def\co{{\cal O}}
\def\n{\nabla}
\def\ci{{\mathcal I}}

\def\<{\langle }
\def\>{\rangle }
\def\[{\left[}
\def\]{\right]}
\def\ch{{\cal H}}

\def\F{{\cal F}}
\newcommand{\bea}{\begin{eqnarray}}
\newcommand{\eea}{\end{eqnarray}}
\newcommand{\be}{\begin{equation}}
\newcommand{\ee}{\end{equation}}
\newcommand{\ba}{\begin{align}}
\newcommand{\ea}{\end{align}}

\renewcommand{\epsilon}{\varepsilon}

   \makeatletter
  \let\over=\@@over \let\overwithdelims=\@@overwithdelims
  \let\atop=\@@atop \let\atopwithdelims=\@@atopwithdelims
  \let\above=\@@above \let\abovewithdelims=\@@abovewithdelims
\renewcommand\section{\@startsection {section}{1}{\z@}%
                                   {-3.5ex \@plus -1ex \@minus -.2ex}
                                   {2.3ex \@plus.2ex}%
                                   {\normalfont\large\bfseries}}

\renewcommand\subsection{\@startsection{subsection}{2}{\z@}%
                                     {-3.25ex\@plus -1ex \@minus -.2ex}%
                                     {1.5ex \@plus .2ex}%
                                     {\normalfont\bfseries}}

\begin{document}
\begin{titlepage}
\unitlength = 1mm
\ \\
\vskip 3cm
\begin{center}

{ \LARGE {\textsc{ Soft Hair on Black Holes}}}

\vspace{0.8cm}
Stephen W. Hawking$^\dagger$, Malcolm J. Perry$^\dagger$ and Andrew Strominger$^*$

\vspace{1cm}
{\it $^\dagger$DAMTP,
Centre for Mathematical Sciences, \\University of Cambridge, Cambridge, CB3 0WA
UK}

{\it $^*$ Center for the Fundamental Laws of Nature,\\ Harvard University,
Cambridge, MA 02138, USA}

\begin{abstract}
It has recently been shown that BMS supertranslation symmetries imply an infinite number of conservation laws for 
all  gravitational theories in asymptotically Minkowskian spacetimes. These laws require black holes to carry a large amount of soft ($i.e.$ zero-energy) supertranslation hair.  The presence of a Maxwell field similarly implies soft electric hair.  This paper gives an explicit description of soft hair in terms of 
soft gravitons or photons on the black hole horizon, and shows that complete information about their quantum state is stored on a holographic plate at the future boundary of the horizon. Charge conservation is used to give an infinite number of exact relations between the evaporation products of black holes which have different soft hair but are otherwise identical. It is further argued that soft hair which is spatially  localized to much less than a Planck length 
cannot be excited in a physically realizable process, giving an effective number of soft degrees of freedom proportional to the horizon area in Planck units. 
 \end{abstract}

\vspace{1.0cm}

\end{center}

\end{titlepage}

\pagestyle{empty}
\pagestyle{plain}

\def\gzz{\gamma_{z\bz}}
\def\gzu{\gamma^{z\bz}}
\def\vx{{\vec x}}
\def\p{\partial}
\def\po{$\cal P_O$}
\def\cN{{\cal N} }
\def\N{${\cal N}~ $}
\def\G{\Gamma}
\def\l{\ell }
\pagenumbering{arabic}

\tableofcontents
\section{Introduction}

Forty years ago, one of the authors argued \cite{swh} that information is destroyed when a black hole is formed and subsequently evaporates \cite{Hawking:1974rv,Hawking:1974sw}. This conclusion seems to follow inescapably from an `unquestionable' set of general assumptions
such as causality, the uncertainty principle and the equivalence principle.  However it leaves us bereft of deterministic laws to describe the universe. This is the infamous information paradox.

Over the intervening years, for a variety of reasons,  the initial conclusion that information is destroyed has become widely regarded as implausible.  Despite this general sentiment, in all this time there has been neither a universally accepted flaw discovered in the  original argument of \cite{swh} nor an a priori reason to doubt any of the `unquestionable' assumptions on which it is based.  

Recently such an a priori reason for doubt has emerged from new discoveries about the infrared structure of quantum gravity  in asymptotically flat spacetimes.  The starting point goes back to the  1962 demonstration by  Bondi, van der Burg, Metzner and Sachs \cite{bms} (BMS) that physical data at future or past null infinity transform non-trivially under, in addition to the expected Poincare transformations, an infinite set of diffeomorphisms known as $supertranslations$. 
These supertranslations separately shift forward or backward in retarded (advanced) time the individual light rays 
comprising future (past) null infinity.  Recently it was shown \cite{Strominger:2013jfa}, using new mathematical results 
\cite{ck} on the structure of null infinity, that a certain antipodal combination of past and future supertranslations is an exact symmetry of gravitational scattering.  The concomitant  infinite number of `supertranslation charge'  conservation laws equate the net 
incoming energy at any angle to the net outgoing energy at the opposing angle. In the quantum theory, matrix elements of the conservation laws give an infinite number of exact relations between scattering amplitudes in quantum gravity. These relations turned out \cite{He:2014laa} to have been previously discovered by Weinberg in 1965 \cite{Weinberg} using Feynman diagrammatics and are known as the soft graviton theorem. The argument may also be run backwards: starting from the soft graviton theorem one may derive both the infinity of conservation laws and supertranslation symmetry of gravitational scattering.

  This exact equivalence has provided fundamentally new perspectives on both BMS symmetry and the soft graviton theorem, as well as more generally the infrared behavior of gravitational theories [5,7,9-39].
 Supertranslations transform the Minkowski vacuum to a physically inequivalent zero-energy vacuum.  Since the vacuum is not invariant, supertranslation symmetry is spontaneously broken.  The soft ($i.e.$ zero-energy) gravitons are the associated Goldstone bosons. The infinity of inequivalent vacua differ from one another by the creation or annihilation of soft gravitons. They all have zero energy but different angular momenta.\footnote{None of these vacua are preferred, and each is annihilated by a different Poincare subgroup of BMS.  
  This is related to the lack of a canonical definition of angular momentum in general relativity.}
  
  Although originating in a different context these observations do provide, as discussed in \cite{Strominger:2014pwa, Pasterski:2015tva},  a priori reasons to doubt  the `unquestionable' assumptions underlying the information paradox:    
  
  \noindent(i) {\it The vacuum in quantum gravity is not unique.} The information loss argument assumes that after the evaporation process is completed, the quantum state settles down to a unique vacuum. In fact, the process of black hole formation/evaporation will generically induce a transition among the infinitely degenerate vacua. In principle, the final vacuum state could be correlated with the thermal Hawking radiation in such a way as to 
  maintain quantum purity. 
  
  \noindent(ii) {\it Black holes have a lush head  of `soft  hair'.} The information loss argument assumes that static black holes are nearly bald: $i.e$ they are characterized solely by their mass $M$, charge $Q$ and angular momentum $J$.  The no-hair theorem \cite{Chrusciel:2012jk} indeed shows that static black holes are characterized by $M$, $Q$ and $J$ up to diffeomorphisms. However BMS transformations are diffeomorphisms which change the physical state. A Lorentz boost for example maps a stationary black hole to an obviously physically inequivalent black hole with different energy and non-zero momentum.  Supertranslations similarly map a stationary black hole to a physically inequivalent one. In the process of Hawking evaporation, supertranslation charge will be radiated through null infinity. Since this charge is conserved, the sum of the black hole and radiated supertranslation charge is fixed at all times.\footnote{In the quantum theory the state will typically not be an eigenstate of the supertranslation charge operator, and the conservation law becomes a statement about matrix elements.} This requires that black holes carry what we call `soft hair' arising from supertranslations.  Moreover, when the black hole has fully evaporated, the net supertranslation charge in the outgoing radiation must be conserved. This will force correlations between the early and late time Hawking radiation, generalizing the correlations enforced by overall energy-momentum conservation. Such correlations are not seen in the usual semiclassical computation. Put another way, the process of black hole formation/evaporation, viewed as a scattering amplitude from $\ci^-$ to $\ci^+$, must be constrained by the soft graviton theorem. 
  
  Of course, finding a flawed assumption underlying the information loss argument  is a far cry from resolving the information paradox. That would require, at a minimum,  a detailed understanding of the information flow out of black holes as well as a derivation of the Hawking-Bekenstein area-entropy law \cite{Hawking:1974rv,Hawking:1974sw,Bekenstein:1973ur}. In this paper we take some steps in that direction.
      
In the same 1965 paper cited above \cite{Weinberg}, Weinberg also proved the `soft photon' theorem. This theorem implies \cite{Strominger:2013lka,He:2014cra,Campiglia:2015qka,Kapec:2015ena,Strominger:2015bla} 
an infinite number of previously unrecognized conserved quantities in all abelian gauge theories - electromagnetic analogs of the supertranslation charges.  By a direct analog of the preceding argument black holes must carry a corresponding `soft electric hair'. The structure in the electromagnetic case is very similar,
but technically simpler, than the gravitational one. In this paper we mainly consider the electromagnetic case,  outlining the gravitational case in the penultimate section. Details of soft supertranslation hair will appear elsewhere.

The problem of black hole information has been fruitfully informed by developments in string theory. In particular it was shown \cite{sv} that certain string-theoretic black holes store complete information about their quantum state in a holographic plate that lives at the horizon. Moreover the storage capacity was found to be precisely the amount predicted by the Hawking-Bekenstein area-entropy law.  Whether or not string theory in some form is  a correct theory of nature, the holographic method it has presented to us of storing information on the black hole horizon is an appealing one, which might be employed by real-world black holes independently of the ultimate status of string theory. 

Indeed in this paper we show that soft hair has a natural description as quantum pixels in a holographic plate. The plate  lives on the two sphere at the future boundary of the horizon. Exciting a pixel corresponds to creating a spatially localized soft graviton or photon on the horizon, and may be implemented by a horizon supertranslation or large gauge transformation.
In a physical setting, the  quantum state of the pixel is transformed whenever a particle crosses the horizon. The combination of the uncertainty principle and cosmic censorship requires all physical particles to be larger than the Planck length, effectively setting a minimum 
spatial size for excitable  pixels. This gives an effective number of soft hairs proportional to the area of the horizon in Planck units and hints at a connection to the area-entropy law. 

It is natural to ask whether or not the supertranslation pixels could conceivably store $all$ of the information that crosses the horizon into the black hole.  We expect the supertranslation hair is too thin to  fully reproduce the area-entropy law. However there are other soft symmetries such as superrotations \cite{deBoer:2003vf,Banks:2003vp,Barnich:2010eb,
  Cachazo:2014fwa, Kapec:2014opa} which lead to thicker kinds of hair as discussed in \cite{Pasterski:2015tva}. Superrotations have not yet been fully studied or understood. It is an open question whether or not the current line of investigation, perhaps with additional new ingredients,  can characterize all the pixels on the holographic plate. 

This paper is organized as follows. Section 2 reviews the analog of BMS symmetries in Maxwell theory, which we refer to as large gauge symmetries. The associated conserved charges and their relation to the soft photon theorem are presented.  In section 3 we construct the extra terms in the conserved charges needed in the presence of a black hole,  and show that they create a soft photon, $i.e.$ excite a quantum of soft electric hair on the horizon. In section 4 we consider evaporating black holes, and present a deterministic formula for the effect of soft hair on the outgoing quantum state at future null infinity. In section 5 we consider physical processes which implant soft hair on a black hole, and argue that a hair much thinner than a planck length cannot be implanted. In section 6 we discuss the gauge dependence of our conclusions. Section 7 presents a few formulae from the generalization from large gauge symmetries and soft photons to BMS supertranslations and soft gravitons. We briefly conclude in Section 8.

\section{Electromagnetic conservation laws and soft symmetries}
In this section we set conventions and review the conservation laws and symmetries of abelian gauge theories in Minkowski space. 

The Minkowski metric in retarded coordinates $(u,r,z, \bz)$ near future null infinity $(\mathcal{I}^+)$ reads
\be\label{ret}
ds^2=-dt^2+ (dx^i)^2=-du^2 -2du dr + 2r^2\gamma_{z\bz}dz d\bz,  
\ee
where $u$ is retarded time and $\gamma_{z\bz}$ is the round metric on the unit radius $S^2$. These are related to standard  Cartesian coordinates by
\be\label{ret}
r^2=x_ix^i, \;\;\; u=t-r, \;\;\; x^i=r\hat{x}^i(z, \bz).  
\ee 
Advanced coordinates $(v,r,z, \bz)$ near past null infinity $(\mathcal{I}^-)$ are\be\label{adv}
ds^2=-dv^2+2dvdr +2r^2\gamma_{z\bz}dz d\bz ;~~~r^2=x_ix^i, \;\;\; v=t+r, \;\;\ x^i=-r\hat{x}^i(z, \bz). 
\ee
$\mathcal{I}^+ $ ($\mathcal{I}^-$) is  the null hypersurface $r= \infty$ in retarded (advanced) coordinates.  
Due to the last minus sign in (\ref{adv}) the angular coordinates on $\mathcal{I}^+$ are antipodally related\footnote{In coordinates with $\gzz=2/(1+z\bz)^{2}$, the antipodal map is 
$z\to-1/\bz$.} to those on $\mathcal{I}^-$ so that a light ray  passing through the interior of Minkowski space reaches  the same value of $z, \bz$ at both $\mathcal{I}^+$ and $\mathcal{I}^-$. We denote the future (past) boundary of $\mathcal{I}^+$ by $\mathcal{I}^+_+$ ($\mathcal{I}^+_-$), and the future (past) boundary of $\mathcal{I}^-$ by $\mathcal{I}^-_{+}$ ($\mathcal{I}^-_{-}$).   
We use conventions for the Maxwell field strength $F=dA$  and charge current one-form $j$ in which $d*F=e^2*j$ (or $\n^aF_{ab}=e^2j_b$) with $e$ the electric charge and $*$ the Hodge dual. 

Conserved charges can be constructed as surface integrals of $*F$ near spatial infinity $i^0$. However care must be exercised as $F$ is discontinuous near $i^0$ and its value depends on the direction from which it is approached. For example, approaching $\mathcal{I}^+_{-}$ from $\mathcal{I}^+$, the radial Lienard-Wiechert electric field for a collection of inertial particles with charges $e_k$ and velocities $\vec v_k$ is, to leading order at large $r$,  
\be F_{rt}={e \over 4\pi r^2}\sum_k {e_k(1-\vec v_k^2) \over (1-\vec v_k \cdot \hat x)^2},\ee
while going to $\mathcal{I}^-_{+}$ from $\mathcal{I}^-$ gives 
\be F_{rt}={e \over 4\pi r^2}\sum_k{e_k (1-\vec v_k^2)\over (1+\vec v_k \cdot \hat x)^2}.\ee
All fields may be expanded in powers of ${1  \over r}$ near $\ci$. We here and hereafter denote the coefficient of ${1 \over r^n}$ by a superscript $(n)$. In coordinates (\ref{ret}),(\ref{adv}), the electric field in general obeys the antipodal matching condition
\be \label{mc} F^{(2)}_{ru}(z,\bz)|_{\mathcal{I}^+_-}=F^{(2)}_{rv}(z,\bz)|_{\mathcal{I}^-_+}.\ee
 (\ref{mc}) is invariant under both $CPT$ and Poincare transformations. 

The matching conditions (\ref{mc}) 
immediately imply an infinite number of conservation laws. For any function $\e(z,\bz)$ on $S^2$ the outgoing and incoming charges defined by\footnote{Here $\e$ is antipodally continuous from  $\mathcal{I}^+_-$ to $\mathcal{I}^-_+$. A second infinity of conserved charges involving the replacement of $F$ with $*F$ \cite{Strominger:2015bla} leads to soft magnetic hair on black holes. This is similar to soft electric hair but will not be discussed herein.}
\bea\label{fr}
	Q_\varepsilon^+ &=& \frac{1}{e^2} \int_{ \mathcal{I}_-^+} \e*F\cr Q_\varepsilon^- &=& \frac{1}{e^2} \int_{ \mathcal{I}_+^-}   \varepsilon  *F \eea
 are conserved:
\be \label{cns} Q^+_\e=Q^-_\e .\ee 
$Q^+_\e$ ($Q^-_\e$) can be written as a volume integral over
 any Cauchy surface ending at $\ci^+_-$ ($\ci^-_+$). In the absence of stable massive particles or black holes, \ip ($\ci^-$) is a Cauchy surface. Hence in this case 
 \bea\label{frb}
	Q_\varepsilon^+ &=&\frac{1}{e^2} \int_{ \mathcal{I}^+} d\e\wedge *F + \int_{ \mathcal{I}^+}\e *j,\cr
Q_\varepsilon^- &=&\frac{1}{e^2} \int_{ \mathcal{I}^-} d\e\wedge *F + \int_{ \mathcal{I}^-}\e *j
 \eea
Here we define $\e$ on all of $\ci$ by the conditions $\p_u\e=0=\p_v\e$.  The first integrals on the right hand sides are zero modes of the field strength $*F$ and hence correspond to soft photons with polarizations $d\e$.   In quantum field theory, (\ref{cns}) is a strong operator equality whose matrix elements are the soft photon theorem \cite{Strominger:2013lka,He:2014cra}. 
The special case $\e=constant$ corresponds to global charge conservation. 

The charges generates a symmetry under which the gauge field $A_z$ on $\ci^+$ transforms as \cite{Strominger:2013lka,He:2014cra}\footnote{ The commutator follows from the standard symplectic two-form $\omega=-{ 1 \over e^2}\int_\Sigma\delta A\wedge *d\delta A$, where $\delta A$ is a one-form on phase space and $\Sigma$ is a Cauchy surface.}
\be \label{sxi} \left[ Q^+_\e ,A_z(u,z,\bz)\right]_{\ci^+}=i\p_z\e(z,\bz).\ee
We refer to this as large gauge symmetry. It is the electromagnetic analog  of BMS supertranslations in gravity. This symmetry is spontaneously broken and the zero modes of $A_z$ -- the soft photons -- are its Goldstone bosons. An infinite family of degenerate vacua are obtained from one another by the creation/annihilation of soft photons. 

\section{Conservation laws in the presence of black holes}
In this section we construct the extra term which must be added to the volume integral expression (\ref{frb}) for the conserved charges (\ref{fr}) in the presence of black holes. 

\begin{figure}
\begin{center}
\includegraphics{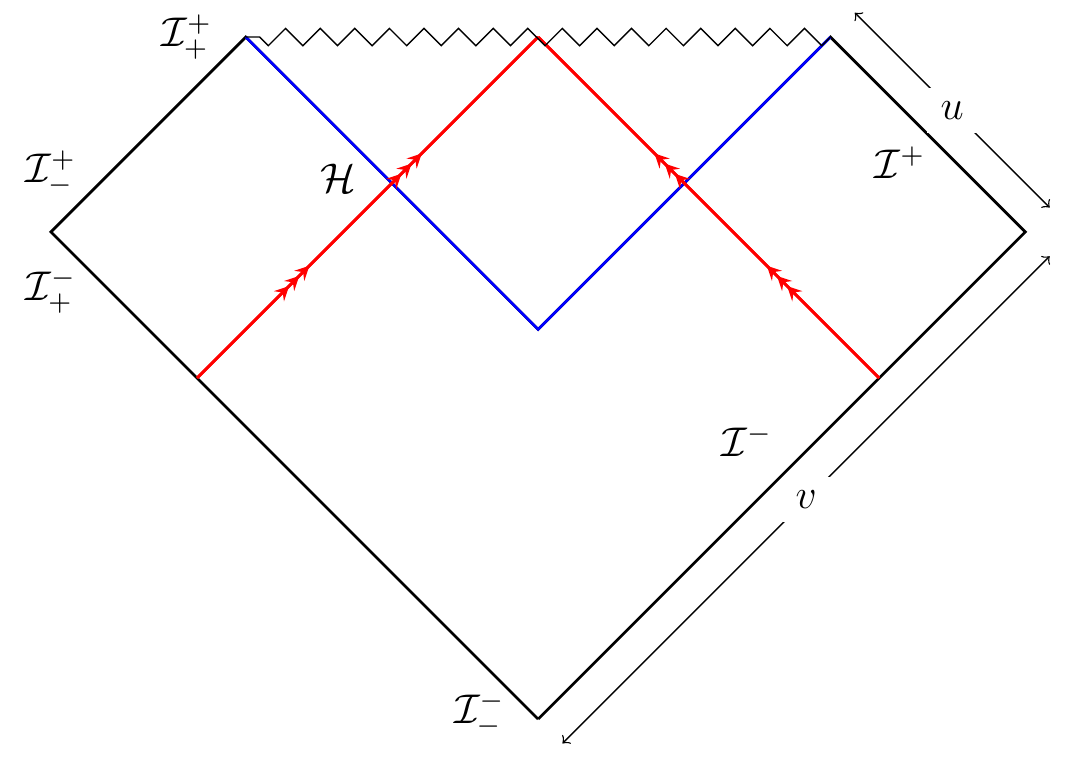}
\end{center}
\caption{Penrose diagram for a black hole formed by gravitational collapse. The blue line is the event horizon. The red line indicates a null spherical shell of collapsing matter. Spacetime is flat prior to collapse and Schwarzschild after. The event horizon $\ch$ and the $S^2$ boundaries $\ci^\pm_\pm$ of $\ci^\pm$ are indicated. $\ci^-$ and $\ci^+\cup\ch$ are Cauchy surfaces for massless fields. } \label{figone}
\end{figure}

When there are stable massive particles or black holes, \ip\ is not a Cauchy surface and a boundary term is needed at $\ci^+_+$ in (\ref{frb}). This problem was addressed for massive particles in \cite{Campiglia:2015qka} where the harmonic gauge condition was used to extend $\e$ into neighborhood of $i^+$ and demonstrate the equivalence of (\ref{cns}) with the soft photon theorem including  massive charged  particles. An alternate  gauge-invariant demonstration was given in \cite{Kapec:2015ena}. In this paper we seek  the extra term for black hole spacetimes.  A simple example of a black hole spacetime, depicted in Figure  (\ref{figone}), is the Vaidya geometry \cite{Vaidya:1951zza} in which a black hole is formed by the collapse of a null shell of neutral matter at advanced time $v=0$:
\be\label{addv}
ds^2=-(1-{2M\Theta(v)\over r})dv^2+2dvdr +2r^2\gamma_{z\bz}dz d\bz,\ee
where $\Theta=0$ ($\Theta=1$) before (after) the shell at $v=0$, and $M$ is the total mass of the shell.  
The event horizon $\ch$ originates at $r=0,~~v=-4M$, continues along $r={v \over 2}+2M$ until $v=0$ after which it remains at  $r=2M$. 
In the absence of massive fields (to which case we restrict for simplicity) which can exit through $i^+$, $\ci^+\cup \ch $ is a Cauchy surface. 
$\ch$ has an $S^2$  boundary in the far future which we denote $\ch^+$. Later we will see $\ch^+$ functions as the holographic plate.  We define a horizon charge as an integral over $\ch^+$ \be \label{rt} Q^\ch_\e={1 \over e^2}\int_{\ch^+}\e*F\ee 
This commutes with the Hamiltonian and hence carries zero energy. 
Extending $\e$ over $\ch$ by taking it to be constant along the null generators, integration by parts yields
\bea Q^\ch_\e={1 \over e^2}\int_\ch d\e\wedge*F+\int_\ch \e*j.\eea
The first term  creates a soft photon on the horizon with  spatial polarization $d\e$. 
$Q^\ch_\e$ generates large gauge transformations on the horizon:
\be\label{sdl} \[ Q^\ch_\e,A_z\]_\ch=i\p_z\e.\ee

At the classical level, the no hair theorem implies for any black hole horizon that $Q^\ch_\e=0$, except for the constant $\l=0$ mode of $\e$ in the case of a charged black hole.\footnote{One may also associate (non-conserved) charges with two-spheres in the horizon other than $\ch^+$. If chosen  during formation or evaporation when the horizon is evolving these charges will in general be nonzero.} One might be tempted to conclude that the $\l>0$-mode charges all act trivially in the quantum theory. 
This would be the wrong conclusion. To see why, lets return momentarily to Minkowski space and consider $Q^+_\e$. Classically in the vacuum there are no electric fields and $Q^+_\e=0$. However neither the electric field nor $Q^+_\e$ vanish as operators. Acting on any vacuum $|0\>$ one has 
\be Q^+_\e|0\>=\left({1 \over e^2}\int_{\ci^+} d\e\wedge*F\right) |0\>\neq 0,\ee
which is a new vacuum with an additional soft photon of polarization $d\e$. Note that the vanishing expectation value $\<0|Q^+_\e|0\>=0$ is consistent with the classical vanishing of the charge. Similar observations pertain to the black hole case. Let $|M\>$ denote the incoming quantum state of a black hole defined on $\ch$. We take it to be formed with neutral matter so that $j=0$ on $\ch$. Then  
\be Q^\ch_\e|M\>=\left({1 \over e^2}\int_H d\e\wedge*F\right) |M\>\neq 0 \ee
is $|M\>$ with an additional soft photon of polarization $d\e$. We refer to this feature which distinguishes stationary black holes of the same mass as soft electric hair. 
\section{Black hole evaporation}
The preceding section gave a  mathematical description of soft electric hair. In this section show that it is physically measurable and therefore not a gauge artefact.

\begin{figure}
\begin{center}
\includegraphics{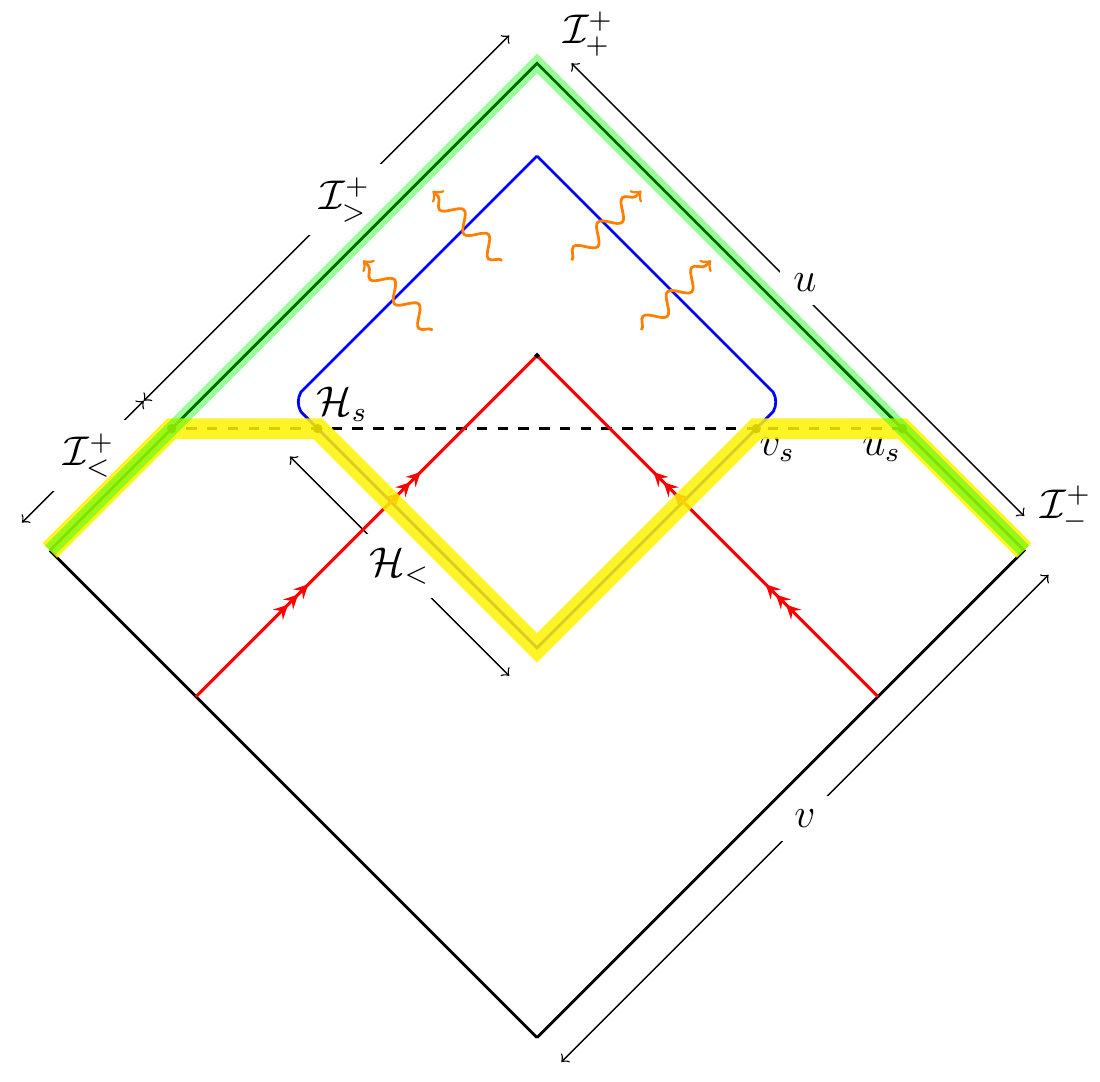}
\end{center}
\caption{Penrose diagram for a semiclassical  evaporating black hole, outlined in blue. The red arrows denote the classical collapsing matter and the orange arrows the outgoing quantum Hawking radiation. The horizontal dashed line divides the spacetime into two regions  long after classical black hole formation is complete and long before the onset of Hawking evaporation. The conserved charges $Q^+_\e$ defined at $\ci^+_-$ can be evaluated as a volume integral either over the green slice comprising $\ci^+$ or the yellow slice involving the classical part of the horizon $\ch_<$. The equality of these two expressions yields infinitely many deterministic constraints on the evaporation process. }\label{figtwo}
\end{figure}

 One way to distinguish the states $|M\>$ and $Q^\ch_\e|M\>$ is to look at their outgoing evaporation products.\footnote{An alternate possibility involves the memory effect. Soft photons exiting at \ip\ can be measured using the electromagnetic memory effect \cite{mema,memb,memc}. The effect falls of like ${1 \over r}$ so the measurement must be done near, but not on, \ip. Similarly we expect a black hole memory effect, enabling the measurement of soft photons on the stretched horizon.} So far we have not let the black holes evaporate.  We work near the semiclassical limit so that the  the formation process is essentially complete  long before the evaporation turns on. We may then divide the spacetime into  two regions via a spacelike splice which intersects \ip\ and the (apparent) horizon at times $u_s$ and $v_s$ long after classical apparent horizon formation is complete but long before Hawking quanta are emitted in appreciable numbers, as depicted in Figure (\ref{figtwo}). We denote by $\ch_<$ the portion of the horizon 
prior to $v_s$  and $\ci^+_<$ (with vacuum $|0_<\>$) and $\ci^+_>$ the portions of $\ci^+$ before and after $u_s$.  We assume that the region between \ip\  and the horizon extending from  $u_s$ to $v_s$ (the horizontal part of the yellow line in Figure (\ref{figtwo})) is empty so that 
$\ci^+_<\cup \ch_<$ is (approximately) a Cauchy slice terminating at $\ci^+_-$. We denote  the two sphere at the future boundary $v=v_s$ of  $\ch_<$ by $\ch_s$.  

For a black hole formed by  a spherically symmetric collapse as in (\ref{addv}), no radiation appears on \ip\ until Hawking radiation turns on after $u_s$.  The incoming quantum state of the black hole, denoted $|M\>$ and defined as a state in the Hilbert space on $\ch_<$, is then a pure state uncorrelated with the vacuum state on $\ci^+_<$.  After a long time evolution, $|M\>$ fully evaporates into some pure (assuming unitarity) outgoing state $|X\>$ of total energy $M$ supported on $\ci^+_>$. Currently, there is no known algorithm for computing $|X\>$, although it is presumably some typical microstate in the thermal ensemble of states produced by the probabilistic Hawking computation. On the other hand, the no-hair theorem, together with causality and a few other basic assumptions, seems to assert that $|X\>$ depends only the total mass $M$ and is insensitive to the details of the quantum state $|M\>$. This is the information paradox. We now show that this assertion fails in a predictable manner due to soft hair.

Consider a second  state 
\be |M' \>= Q_\e^{\ch_<}|M\>,\ee
where  $Q_\e^{\ch_<}={1 \over e^2}\int _{\ch_s}\e*F$. 
Since no charges were involved in the formation of $|M\>$, $j|_\ch$ vanishes and the action  of $Q_\e^{\ch_<}$ creates a soft photon on the horizon.  Hence $|M'\>$ and  $|M\>$ differ only by  a soft photon and  are energetically degenerate.\footnote{ Up to corrections of order ${1 \over u_s}$, which we assume negligible. The two states do carry different angular momentum. This however is not the primary origin of their difference: in a more complicated example, we might  have added several soft photons with zero net angular momentum and still been able to distinguish the resulting final state.} 
$|M'\>$ eventually evaporates to some final state which we denote $|X'\>$ supported on $\ci^+_>$. 

We wish to use charge conservation to relate the two outgoing states   $|X'\>$ and $|X\>$.
Since, in the very far future there is nothing at $i^+$ we may write (using the green Cauchy surface in Figure(\ref{figtwo}))
\be\label{ra} Q_\e^+=Q_\e^{\ci^+}=Q_\e^{\ci^+_<}+Q_\e^{\ci^+_>},\ee
where the last two terms denote volume integrals 
\be Q_\varepsilon^{\ci^+_{>,<}} =\frac{1}{e^2} \int_{ {\ci^+_{>,<}} } d\e\wedge *F + \int_{ {\ci^+_{>,<}} }\e *j.\ee
On the other hand, approximating  the yellow Cauchy surface in Figure (\ref{figtwo}) by  $\ci^+_<\cup \ch_<$ we may also write 
\be \label{rb}Q_\e^+=Q_\e^{\ci^+_<}+Q_\e^{\ch_<}.\ee
Comparing (\ref{ra}) and (\ref{rb}) it follows that\footnote{This relation is similar in spirit to some that appear in \cite{'tHooft:1996tq,Polchinski:2015ce}. Yet it differs in detail: for example the effect in \cite{'tHooft:1996tq,Polchinski:2015ce} is largest for equal incoming and outgoing angles, in our case  (see the next section) the effect is centered around the antipodal outgoing angle. Perhaps future work will relate these effects. We thank Joe Polchinski for discussions on this point. }

\be |X' \>= Q_\e^{\ci^+_>}|X\>.\ee
We note that the action of $Q_\e^{\ci^+_>}$ on $|X\>$ involves both hard and soft pieces and can be quite nontrivial. 

    In writing (\ref{rb}) we use the gauge parameter $\e$ defined on $\ch$ in equation (\ref{rt}) in the coordinates (\ref{addv}). In principle,  as was demonstrated for the analysis of massive particles at \ip\ \cite{Kapec:2015ena},  physical observables should not depend on how the gauge parameter is extended into the interior from $\ci^+$. This is discussed in section 6 below where some consistency checks for gauge invariance are given. 
    
    In this section we have made several simplifying assumptions and approximations including that classical black hole formation and quantum evaporation can be temporally separated, that $\ci^+_<\cup\ch_<$ is a Cauchy surface and that the collapse is spherically symmetric. It is important to stress that the 
 existence of an infinite number of deterministic relations between incoming black hole states and their outgoing evaporation products is independent of these restrictions and assumptions. It follows solely from the exact equality of the two expressions for the charge as volume integrals on the early (yellow) and late (green) Cauchy slices. The purpose of the approximations and assumptions was simply to find a context in which a concise and simple statement of the consequences of charge conservation could be made. 

In summary, once we have determined the outgoing state arising from $|M\>$, we can predict the exact outgoing state arising from the action of a large gauge transformation of  $|M\>$, even though the resulting state is energetically degenerate. This contradicts the assertion that the outgoing state depends only on the total mass $M$.  The assertion fails because it was based on the quantum-mechanically false assumption that  black holes have no hair. This is the content of the statement that quantum black holes carry  soft electric hair. 
\section{Quantum hair implants}
In the preceding section we demonstrated that black holes with different numbers of horizon soft photons
are distinguishable. In this section we  show that the soft photon modes on the horizon can be indeed excited in a physically realizable process, as long as their spatial extent is larger than the Planck length $L_p$.

 Soft photons at \ip\  are excited whenever charge crosses \ip\  with an
  $\l>0$ angular momentum profile. 
Similarly, soft photons on the horizon are excited whenever charge is thrown into the black hole with an 
  $\l>0$ angular momentum profile. To see this, consider a null shock wave thrown into the black hole geometry (\ref{addv}) at $v=v_0>0$, with divergence-free charge current in an angular momentum eigenstate 
\be \label{dti} j_v={Y_{\l m}(z,\bz)\over r^2}\delta (v-v_0)\ee 
with $\l >0$ and $Y_{\l m}$ the usual spherical harmonics. 
We neglect the backreaction of the shell and consequent electromagnetic field on the geometry.
The leading large-$r$ constraint equation for the Maxwell field on $\ci^-$ may then be written  \be \p_vF^{(2)}_{rv}+\gzu( \p_zF^\0_{\bz v}+\p_\bz F^\0_{zv})=e^2 j^\2_v.\ee We wish to consider initial data with no photons - hard or soft - which means $F^\0_{zv}=0$.  This implies the ${1 \over r^2}$ coefficient of the electric field is 
\be \label{reg} F^{(2)}_{rv}=e^2 Y_{\l m}\Theta(v-v_0).\ee In general the shell  will produce radiation into both $\ci^+$ and $\ch$.  The no-hair theorem implies that in the far future of both $\ci^+$ and $\ch$ \be
F_{vr}|_{\ch^+}=F^{(2)}_{ur}|_{\ci^+_+}=0.\ee Integrating the constraints on $\ci^+$ and using the matching condition (\ref{mc}) then implies that 
\be\label{ras} \p_z\int _{-\infty}^\infty dv F^\0_{\bz v }+\p_\bz \int _{-\infty}^\infty dvF^\0_{zv} =\gzz e^2 Y_{\l m}. \ee
The solution of this is 
\be \int _{-\infty}^\infty dvF^\0_{zv}= -{e^2\over \l(\l+1)} \p_z Y_{\l m}, \ee
or equivalently in form notation 
\be \int_{\ci^+} F\wedge \hat \star d Y_{\l' m'} ={e^2} \delta_{\l\l'}\delta_{mm'},\ee
where $\hat \star$ is the Hodge dual on $S^2$. 
This \ip\ zero mode of $F^\0_{zv}$ corresponds to a soft photon with polarization vector proportional to $ \p_z Y_{\l m}$. 
Similarly integrating $d*F=e^2*j$ over $\ch$ with $j$ given by (\ref{dti}) yields 
\be \int_{\ch} F\wedge \hat \star d Y_{\l' m'} ={e^2} \delta_{\l \l'}\delta_{m m'}.\ee
This is a soft photon on the horizon with polarization vector proportional to $ \p_z Y_{\l m}$. It is created by the soft part of the charge operator $Q^\ch_{\e_{\l m}}$ with 
\be \e_{\l m}= -{e^2 \over \l(\l+1)}  Y_{\l m}.\ee

It is interesting to note that not all horizon soft photons can be excited in this manner. 
Suppose we want to excite a soft photon whose wave function is localized within as small as possible a region of area
$L^2$ on the black hole horizon. Then we must send in a charged particle whose cross-sectional size is $L$ as it crosses the  horizon.  However, no particle can be localized in a region smaller than either its Compton wavelength $\hbar \over M$ or its Schwarzschild radius $ML_p^2$. The best we can do is to send in a small charged black hole of Planck size $L_p$ and Planck mass $M_p$. This will excite a soft photon of spatial extent $L_p$ (or larger) on the horizon.  We cannot  excite soft photons  whose size is parametrically smaller than the Planck length. It follows that the effective number 
of soft photon degrees of freedom is of order the horizon area in Planck units. The entropy of such modes is naturally of order the area, a tantalizing hint of a connection to the Hawking-Bekenstein area-entropy law.

\section{Gauge invariance}

In this section we consider the relation of the gauge parameters on $\ci$ and $\ch$. 
In principle no physical conclusions should depend on how we continue large gauge transformations at $\ci$  to the horizon. In the above we used  $\e|_{\ch}= \e|_{\ci^-}$ in advanced coordinates (\ref{addv}). Suppose we instead took 
\be \e|_{\ch}=\alpha \e|_{\ci^-}, \ee
for some constant $\alpha$. Then we would have 
\be \label{pop}Q_\e^+=Q_\e^{\ci^+}+Q_{\alpha \e}^\ch.\ee On the other hand  the soft photon we implanted  in the  preceding section  would be created by the action of  $ Q^\ch_{\e_{\l m}/ \alpha}$ instead of just $Q^\ch_{\e_{\l m}}$.  Repeating the argument of section 4 using (\ref{pop}), one would then find that the hair implant would affect the final state of the  black hole evaporation by the action of $Q^{\ci^+_>}_{\e_{\l m}}$. Hence the factors of $\alpha$ cancel as required by gauge invariance. These conclusions would remain even if $ \e|_{\rm horizon}$ were a nonlocal or angular-momentum dependent function of $\e|_{\ci^-}$. For example we might have taken $\e|_{\ch}= \e|_{\ci^+}$ using $\alpha \sim(-)^\l$. 
However the chosen identification 
\be\label{dsa} \e|_{\ch}=\e|_{\ci^-} \ee
in advanced coordinates is natural because excitations of the horizon are naturally sent in from $\ci^-$, and large gauge transformations on $\ch$ are simply connected to those on $\ci^-$. 
To see this consider the charged shock wave, but instead of solving the constraints with a radial electric field as in (\ref{reg}), let us instead take
\be F_{vz}=\p_z \e_{\l m} \delta(v-v_0),~~F_{vr}=0.\ee
Then the electromagnetic field is in the vacuum both before and after the shock wave. However, the gauge potential must shift. In temporal gauge $A_v=0$ we have
\be A_z=\Theta(v-v_0) \p_z\e_{\l m}.\ee
This equation is valid on both $\ci^-$ and $\ch$. 
The shock wave is a domain wall which interpolates between two vacua which differ by a large gauge transformation parameterized by $\e_{\l m}$.\footnote{A discussion of domain walls interpolating bewteen BMS-inequivalent vacua, and their relation to gravitational memory,  was given in \cite{Strominger:2014pwa}.}  In this context $\e_{\l m}$ naturally obeys (\ref{dsa}).  
\section{Supertranslations}
The situation for BMS supertranslations is very similar. We give an overview here, details will appear elsewhere.

BMS supertranslations \cite{bms} are diffeomorphisms which act infinitesimally near  $\ci^+$ as 
\be \zeta_f =f\p_u -{ \gzu \over r}(\p_z f\p_\bz+\p_\bz f \p_z)+....\ee
where $f(z,\bz)$ is any function on $S^2$. The indicated corrections are further  subleading in ${1 \over r}$ and depend on the gauge choice. The extension of this to  $\ci^-$ obeying 
\be \zeta_f =f\p_v +{ \gzu \over r}(\p_z f\p_\bz+\p_\bz f \p_z)....\ee
where, as in (\ref{ret}),(\ref{adv}), $z$ near $\ci^-$ is antipodally related to $z$ on \ip.

The metric in the neighborhood of the horizon $\ch$ may always be written \cite{Lee:1990nz}
\be \label{hm}ds^2= 2dvdr+g_{AB}dx^Adx^B +\co(r-r_\ch),\ee
with the horizon located at 
at $r=r_\ch$, $v$ an (advanced) null coordinate on $\ch$,  $r$ an affine ingoing null coordinate and $x^A,~A,B=1,2$ spatial coordinates on $\ch$.  We define horizon supertranslations by
\be \label{hst} \zeta=f\p_v-(r-r_\ch)g^{AB}\p_A f\p_B ...\ee
where here the gauge dependent corrections are suppressed by further powers of $(r-r_\ch)$. These diffeomorphisms preserve the form (\ref{hm}) of the metric. 
Horizon supertranslations have been studied from a variety of viewpoints in \cite{Koga,Hotta:2000gx,Blau:2015nee}. Some of these authors allow $f$ to depend on $v$. However, as we saw in the the Maxwell case, the charge can be written as a surface integral on the future boundary $\ch^+$ of $\ch$. This implies that - in the context we consider - the charges associated to the extra symmetries associated with $v$-dependent $f$s vanish identically and are trivial even in the quantum theory.  So for our purposes it suffices to restrict this freedom and impose $\p_vf=0$.  We note that the ingoing expansion on $\ch$, $\theta_r=\half g^{AB}\p_r g_{AB}$, transforms inhomogeneously under supertranslations:
\be \delta_f \theta_r|_\ch= -D^2f,\ee
where $D^2$ is the laplacian on the horizon. 
Hence $\theta_r$ may be viewed as  the Goldstone boson of spontaneously broken supertranslation invariance.

We  wish to compute the general-relativistic symplectic structure  $\omega$  of linearized metric variations on the solution space with horizon supertranslations (\ref{hst}). The general expression for two metric variations $h_{ab}$ and $h'_{cd}$  around a fixed background metric $g_{ef}$ is \cite{Ashtekar:1981bq,Crnkovic:1986ex,wz} \be \omega(h,h^\prime) = \int_{\Sigma} *J (h,h^\prime) \ee
where $\Sigma$ is a Cauchy  surface and $J=J_a dx^a$ is the symplectic one-form. Naively $\omega$ should vanish if either $h$ or $h'$ are pure gauge. However this is not quite the case 
if the diffeomorphism $\zeta$ does not vanish at the boundary of $\Sigma$. One finds \be *J(h,\zeta)=-{1 \over 16\pi G}d*\F,\ee
where 
\be \label{dcz}\F = \frac{1}{2}hd\zeta-dh\wedge \zeta
+ \left(\zeta_c\n_ah^c{}_b 
+ h^c{}_b \n_c\zeta_a\ 
-\zeta_a\n_c h^c{}_b\ \right) dx^a\wedge dx^b.   \ee
This gives the symplectic form 
\be\label{sfm} \omega(h,\zeta)=-{1 \over 16 \pi G}\int_{\p \Sigma} *\F.\ee
Wald and Zoupas \cite{wz} derive from this a formula for the differential charge associated to the  diffeomorphism $\zeta$, which for supertranslations is simply integrated to 
$Q_f=\omega(g, \zeta_f).$  For $\ci^+_-$ this reproduces the familiar formula in Bondi coordinates 
$Q_f^+= {1 \over 4\pi G}\int_{\ci^+_-}d^2z\gzz f m_B$ with $m_B$ the Bondi mass aspect. For $\Sigma=\ch$ and $\zeta$ the horizon supertranslation (\ref{hst}), this reduces to 
\be \label{swd} Q^\ch_f= -{1 \over 16\pi G}\int_{\ch^+}d^2x\sqrt{g} fg^{AB}\p_v h_{AB}.\ee

$Q^\ch_f$, like $Q^\ch_\e$,  vanishes for all stationary classical solutions \cite{Flanagan:2015pxa}.\footnote{This point needs further clarification. The proof in \cite{Flanagan:2015pxa} assumed time-independence of all metric components in Bondi coordinates. It is plausible that all solutions asymptote at late times to such geometries, up to a possibly non-trivial diffeomorphism, but we do not know a proof of this.} Hence stationary  black holes do not carry classical supertranslation hair, just as they do not carry classical electric hair. However, as in the electric case, the action of $Q^\ch_f$ 
creates soft gravitons on the horizon.  From this point forward, the argument that black holes carry soft supertranslation hair proceeds in a nearly identical fashion to the electromagnetic case.

\section{Conclusion}

We have reconsidered the  black hole information paradox in light of recent insights into the infrared structure of quantum gravity. An explicit description has been given of a few of the pixels in the holographic plate at the future boundary of the horizon.  Some information is accessibly stored on these pixels in the form of soft photons and gravitons. A complete description of the holographic plate and resolution of the information paradox remains an open challenge, which we have presented new and concrete tools to address.

\bigskip

\centerline{\bf Acknowledgements}
We are grateful to Gary Gibbons, Sasha Hajnal-Corob, Dan Harlow, Prahar Mitra, Anjalika Nande, Joe Polchinski, Ana Raclariu and Sasha Zhiboedov.
This work was supported in part by the STFC, NSF grant 1205550, the Cynthia and George Mitchell Foundation and a Bershadsky Visiting Fellowship to MJP.

\end{document}